# Agile requirements engineering practices: a survey in Brazilian software development companies

Juan Carlos Barata[1], Diego Lisboa[2], Laudelino Cordeiro Bastos[1], Adolfo Neto[1]

[1] Universidade Tecnológica Federal do Paraná (UTFPR), Curitiba, Brazil

`juan.barata.si@gmail.com, bastos@utfpr.edu.br, adolfo@utfpr.edu.br`

[2] Universidade Federal do Pará (UFPA), Belém, Brazil

`diegolisboa@ufpa.br`

**Abstract.** Requirements Engineering (RE) is one of the prime areas in software development. Since agile software development englobes several emerging techniques and advocates for continuous improvement, it urges the question of which agile RE practices are currently most used, their characteristics, and the challenges in their employment. The aim of this work is to investigate and categorize the collection and specification of agile requirements practices based on how professionals perceive their importance for a software project that applies agile methodologies. Thus, a survey was carried out with forty-six (46) Brazilian software development professionals, inquiring which methods are used for the collection and specification of agile requirements, as well as the features, benefits, and difficulties when employing the methods. The responses allowed us to perform data analysis and identify the relationships between the respondents' experience and the viewpoints on the collection methods and the agile requirements specification. In addition, it was noted that the adoption of these methods is still very recent. They have mainly been used for less than five years. Moreover, it was noted that, for most respondents, there are yet significant challenges and advances to be made for better efficiency in applying the informed methods.

## 1       Introduction

The main objective of requirements engineering (RE) is to identify the demands of stakeholders, which are people or organizations that will be affected by the system and possess influence, direct or indirect, over the system requirements. Consequently, it is essential to understand the issue and its context, elicit the requirements for the system, analyze, document, and validate them [14].

RE is intended to elicit, organize, and document the software requirements based on a process that establishes and maintains an agreement among stakeholders [8]. It is a communication process that will stipulate what the software must do, its functions, essential and desirable properties, and restrictions [23].

RE practices employed by agile teams differ from those applied in traditional RE [5]. Similar to any process, RE must be continuously improved. Agile software development methods recommend adopting a step-by-step approach to improve software development processes and adaptation to the altered conditions (e.g., focusing on introducing or improving unique practices) rather than changing all at once.

Over the years, many survey studies conducted have investigated the selection level of agile practices [1, 2, 4, 6, 7, 9, 15, 17, 18, 19, 21] in software development projects, including the 'State of Agile' report (Digital.ai, 2020). Therefore, it is apparent the importance of agile methods and their practices for the software development industry.

The research question in this study is based on the following gap: "What are the main techniques, characteristics, and challenges of employing agile methods in the collection and specification of requirements?". We described the employment of the collection and specification of agile requirements adopted by software development companies, focusing on techniques, characteristics, and challenges of the technique practitioners. A survey was carried out, where forty-six (46) participants and practitioners from thirty-one (31) Brazilian companies (public and private) answered about the numerous collection and specification techniques of agile requirements. The results exhibited an overview of the employment of these techniques in private and public organizations.

The remainder of this work is organized as follows: Section 2 presents the description of the work reports; Section 3 displays the research methodology; Section 4 shows the research report; Section 5 presents the discussion and conclusion of the study.

## 2    State of the Art

Recently, the literature has described the employment of RE in projects that adopt agile methods and demonstrated their importance since one of the main reasons for software projects failures is the poor collection of requirements [11]. According to SAITO [20], a successful software project depends on the quality of the Software Requirements Specification (SRS) model. SRS inadequately done is a catalyst for other issues throughout the software project, mainly when the process adopts agile forms of software development, as many practitioners skip some steps and produce important artifacts, causing inefficiency in the requirements specification.

In the work published by MEDEIROS [16], the author emphasizes that an inadequate SRS is an enhancer of other issues throughout the software project, especially when applying agile methods. According to the author, it was indicated that in agile environments, SRS are generally superficial, insufficient, and inadequate. Consequently, the author proposed an agile approach for SRS, using a set of good practices from Agile-RE, such as Agile Modeling, Prototypes, and Scenario Testing features, in order to propose an agile SRS process.

Thus, by examining several techniques and approaches for the collection and specification of requirements in agile software development projects, this work aims to investigate the techniques, characteristics, and challenges in the collection and specification of agile requirements currently employed by private and public companies through a questionnaire, to collect accurate data on the full adoption of agile methods in software development.

## 3    Research Methodology

This study aims to illustrate the characteristics and challenges in employing the collection and specification of agile requirements by private and public companies. A survey was carried out as the research method. According to WOHLIN [25], surveys are used when a technique or tool has already occurred or before it was conducted.

The characteristics are described for which agile method is employed, how the respondents first learned about Agile-RE, level of knowledge, time of use, and benefits achieved after applying the method. The challenges were defined through hypothesis and determined according to the authors cited in Section 2. Overall, a total of five (5) hypotheses were established, which are listed below:

- H1. Low client availability is a relevant challenge;
- H2. Inadequate c between client and team is a relevant challenge;
- H3. The lack of transparency between client's needs and solutions is a relevant challenge;
- H4. Inefficient change control in requirements is a relevant challenge;
- H5. Insufficient documentation for implementation, maintenance, and/or training is a relevant challenge.

### 3.1    Survey validation

Before sharing, the survey was validated with the research group: a professor from Universidade Tecnológica Federal do Paraná (UTFPR), a professor from Universidade Federal do Pará (UFPA), a part of the software development team from Coordenadoria de Tecnologia da Informação e Comunicação da Pró-Reitoria de Ensino de Graduação (COTIC/PROEG) da UFPA and a part of the software development team from a private company. The questions from the pilot survey were answered by the research group and, after that, the survey feedback was made.

### 3.2 Planning and scheduling the survey

The survey was conducted virtually through the Google Forms platform. The link for the survey was shared in five social media: LinkedIn, Facebook, Instagram, WhatsApp, and Telegram. The collected data was stored in the Zenodo online repository and is available in the Portuguese version.

### 3.3 Analyzing the results

The responses were analyzed through frequency and percentages that were automatically generated by the Google Forms platform.

## 4 Results

A total of forty-six (46) responses were collected, in which forty-four (44) respondents informed that they work with collection and specification of requirements, and only two respondents do not work with requirements but are a part of a software development team. 69.6% of respondents work in private companies and 30.4% in public companies. 91.7% of respondents work in national companies and 8.3% in multinational companies, working in different positions, as shown in Figure 1.

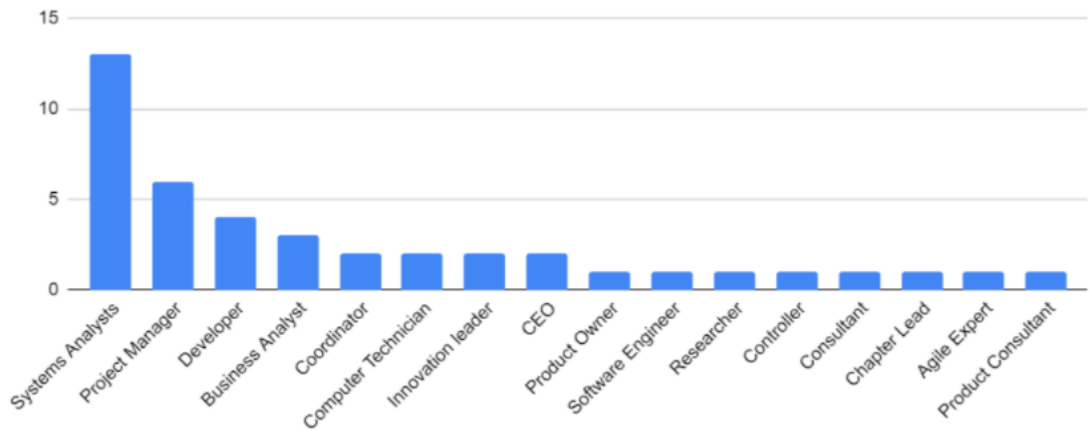

**Fig. 1.** Distribution of working positions among the survey respondents.

### 4.1 Characterization of employed methods for the collection of requirements

In the first question of the questionnaire, we identified which methods are employed by companies for the collection of requirements by inquiring: "What is the main method used for the collection of requirements in software projects that employ agile methods?". Table 1 shows that Interviews with Stakeholders (56.6%), Lean Inception (17.4%), and Design Thinking (13%) are the most used methods.

**Table 1.** Percentage of respondents that informed the use of each collection of requirements method.

| Collection of requirements methods | Percentage of respondents (%) |
| --- | --- |
| Interview with stakeholders | 56.6 |
| Lean Inception | 17.4 |
| Design Thinking | 13 |
| Workshops | 4.3 |
| Lean Iron Analysis | 2.2 |
| Others | 2.2 |

Based on the results, 4.4% of respondents stated that they learned about the method through internet research, 30.4% learned through the academy, 23.9% learned through training or courses, and 39.2% learned through their workplace.

As reported in the survey, 10.9% of respondents claimed that collection of requirements is used in their team for less than one (1) year, 23.9% use it between one (1) and two (2) years, 23.9% use it between two (2) and four (4) years, 8.7% use it between four (4) and five (5) years, and 32.6% use it for more than five (5) years. Moreover, 2.2% of respondents informed that they possess little knowledge about the method, 36.9% have intermediate knowledge, 28.3% have high knowledge, and 32.6% have enough knowledge about the method to consider themselves experts.

In the survey, we asked: "What would be the benefits achieved by adopting the collection of requirements methods employed in the company they operate?". Figure 2 exhibits that the main benefits achieved were: Understanding the client's needs, Scope visibility, Faster validation, MVP (Minimum Viable Product) alignment, and team collaboration. The least mentioned benefits were: Practicality, Standardization, and Satisfaction. In addition, three (3) respondents were unable to respond to the question. The respondents could mention more than one benefit.

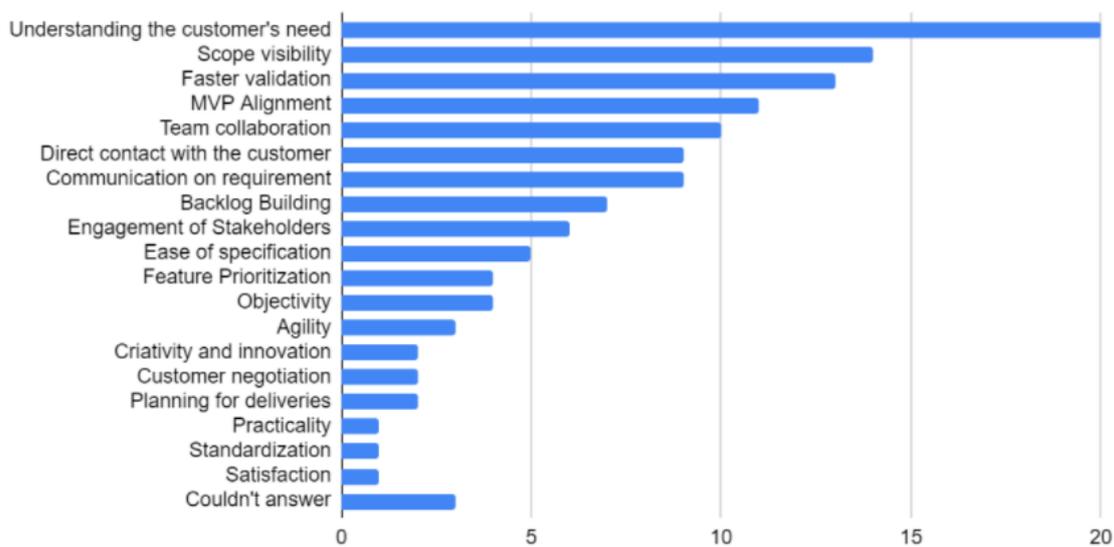

**Fig. 2.** Benefits of employing the collection of requirements methods mentioned by the respondents.

### 4.2     Characterization of employed methods for the specification of requirements

After finishing the questions about the collection of requirements, we seek to identify which methods are used for the specification of requirements. We asked: "What are the main methods employed for the specification of requirements in software projects that uses agile methods?". Table 2 displays the most used methods: Users History with 60.9%, Product Backlog Building (PBB) with 15.2%, and Prototypes with 13.0%.

**Table 2.** Percentage of respondents that informed the use of each specification of requirements method.

| Specification of requirements methods | Percentage of respondents (%) |
|---|---|
| Users History | 60.9 |
| Product Backlog Building (PBB) | 15.2 |
| Prototypes | 13.0 |
| Use Cases | 10.9 |

Based on the results, 2.2% of respondents stated that they learned about the method through internet research, 21.8% learned through the academy, 41.3% learned through training or courses, and 34.8% learned through their workplace.

Concerning the time of use, 19.6% stated that specification of requirements is used in their team for less than one (1) year, 21.7% use it between one (1) and two (2) years, 32.6% use it between two (2) and four (4) years, 2.2% use it between four (4) and five (5) years, and 23.9% use it for more than five (5) years. Furthermore, 2.2% of respondents informed that they possess little knowledge about the method, 36.9% have intermediate knowledge, 32.6% have high knowledge, and 28.3% have enough knowledge about the method to consider themselves experts.

We also asked: "What would be the benefits achieved by adopting the specification of requirements methods employed in the company they operate?". Figure 3 reveals that the main benefits cited were: Clarity, Problem identification, Objectivity, Features Partitioning, and Features Listing. Adherence to the process, The use of documentation, and Standardization were the least mentioned benefits. The respondents could mention more than one benefit.

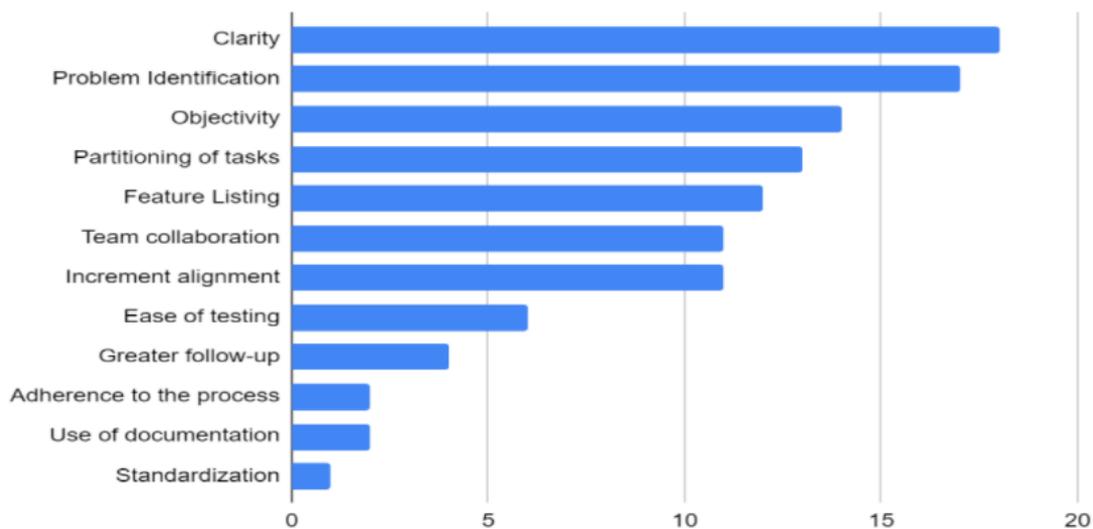

**Fig. 3.** Benefits of employing the specification of requirements methods mentioned by the respondents.

The questionnaire also included questions about the work team satisfaction with the employment of the collection and specification of requirements methods mentioned by the respondents. For 19.6% of respondents, the team is neither satisfied nor unsatisfied, 50% is more or less satisfied, and 30.4% is very satisfied.

### 4.3  Challenges in the employment of the collection and specification of requirements methods

In order to verify and identify the challenges in the employment of the collection and specification of requirements methods, we asked: "What are the difficulties, limitations, challenges, and points to improve in the employment of the collection and specification of requirements methods?". 73.9% of respondents informed "Yes" and 26.1% informed "No". It is essential to highlight that the five (5) hypotheses of possible challenges were added to the question to verify their relevance. Table 3 shows the indication percentage that respondents marked as a challenge in employing the methods.

**Table 3.** Challenges in adopting the collection and specification of requirements mentioned by the respondents.

| Challenges in adopting the collection and specification of requirements | Percentage of respondents (%) |
|---|---|
| Low client availability | 39.5 |
| Inefficient change control in requirements | 18.4 |
| Insufficient documentation | 13.2 |
| Inadequate interaction | 13.2 |
| The lack of transparency between client's needs and solution | 5.3 |
| Lack of knowledge from the development team | 2.6 |
| Lack of domain and agile mindset from the team | 2.6 |
| Platform limitations | 2.6 |
| Lack of knowledge about the methods | 2.6 |

Concerning the initial hypotheses, the following results were obtained:

- H1. Low client availability is a relevant challenge: according to Table 4, it was identified that low client availability possesses the highest indication percentage (39.5%);
- H2. Inadequate interaction between client and team is a relevant challenge: it was noted that inadequate interaction between client and team had an indication percentage of 13.2%. The same percentage was observed in H5;
- H3. The lack of transparency between client's needs and solutions is a relevant challenge: it was observed that the lack of transparency between client's needs and solutions was the least indicated by respondents (5.3%);
- H4. Inefficient change control in requirements is a relevant challenge: H4 was the second most indicated challenge with a percentage of 18.4%;
- H5. Insufficient documentation for implementation, maintenance, and/or training is a relevant challenge: The percentage observed in H2 was the same as H5, with 13.2%.

## 5  Conclusions

This work aimed to describe the characteristics and challenges in employing the collection and specification of agile requirements by private and public companies. A survey was carried out as the research method, where forty-six (46) responses were collected from practitioners in software analysis and development teams. The results showed the characteristics and challenges of the collection and specification of requirements methods adopted by private and public companies when using agile software development methods.

In projects that employ agile methods, when analyzing the collection of requirements, it was observed that the main applied methods are Interviews with Stakeholders, Lean Inception, and Design Thinking. In addition, it was noticed that these methods are being used for less than five (5) years (67.4% of respondents). However, 32.6% of respondents stated that they know enough about the methods to consider themselves experts. The main benefits in using the collection of requirements methods mentioned by the respondents are: Understanding the client's needs, Scope visibility, and Faster validation.

In relation to the specification of requirements in projects that employ agile methods, the Users History, PBB, and Prototypes were the most mentioned methods used. It was also noticed that 73.9% of respondents use these requirements methods for less than four (4) years, and 97.8% consider themselves with intermediate, high, and advanced knowledge about the methods. According to the respondents, the main benefits in employing the specification of requirements methods are: Clarity, Problem identification, and Objectivity.

Concerning the challenges and points to improve in the employment of collection and specification of agile requirements, 73.9% of respondents informed that there are challenges and points of improvement. The four (4) main relevant challenges identified were: low client availability, inefficient change control in requirements, insufficient documentation for implementation, maintenance, and/or training is a relevant challenge, and inadequate interaction between client and team.

Therefore, the collected data suggest that the employment of collection and specification of agile requirements are still very recent and are mainly used for less than five (5) years, but there are several knowledgeable workers that can apply these methods accordingly. Nevertheless, for most respondents, the collection and specification of requirements still have significant challenges and improvements for their proper and efficient use.

Overall, the data allows new research options for the future, such as "what strategies could be applied to obtain more availability from clients?", "what can be done to maintain an efficient change control in requirements?", "how to maintain and improve the interaction with stakeholders?", and "how to improve projects documentation?". These research options can assist in improving the employment of Agile-RE in software analysis and development project teams in private and public organizations.

# References

1. Ali, M.A., 2012. Survey on the state of agile practices implementation in Pakistan. Int. J. Inf. Commun. Technol. Res. 2 (4).
2. Barabino, G., Grechi, D. Tigano, D., Corona, E., Concas, G., 2014. Agile methodologies in web programming: a survey. Proc. Of Intl Conference on Agile Software Development. Springer.
3. Boness, K.D., Harrison, R., 2007. Goal sketching: towards ágil requirements engineering. In: second International Conference on Software Engineering Advances (ICSEA).
4. Buchalcevova, A., 2009. Research of the use of agile methologies in the Cezch Replubic. Information Systems Development. Springer.
5. Cao, L., Ramesh, B., 2008. Agile requirements engineering practices: an empirical study. Softw., IEEE 25(1).
6. Causevic, A., Sundmark, D., Punnekkat, S., 2010. An industrial survey on contemporary aspects of software testing. Proc. Of 3rd Intl Conference on Software Testing, Verification and Validation (ICST). IEEE.
7. Doyle, M., Williams, L., Cohn, M., Rubin, K.S., 2014. Agile software development in practice. Proc. Of Intl Conference on Agile Software Development. Springer.
8. Engholm, H., 2010. Engenharia de Software na Prática, São Paulo: Novatec.
9. Hussain, Z., Slany, W., Holzinger, A., 2009. Current state of agile user-centered design: a survey. Proc. Of 5th Symposium of the Workgroup HCI and Usability Engineering of Austrian Computer Society on HCI and Usability for e-Inclusion. Springer.
10. Inayat, I., Salim, S.S., Marczak, S., Daneva, M., Shamshirband, S., 2015. A systematic literature review on agile requirements engineering practices and challenges. Compt. Human Behav. 51.
11. Kassab, M., 2015. The changing landscape of requirements engineering practices over the past decade. Proc. of 5th Intl Workshop on Empirical Requirements Engineering (EmpiRE). IEEE.
12. Kassab, M., 2014. An empirical study on the requirements engineering practices for agile software development. Proc. of 40th EUROMICRO Conference on Software Engineering and Advanced Applications (SEAA). IEEE.
13. Kassab, M., Neill, C., Laplante, P., 2014. State of practice in requirements engineering: contemporary data. Innov. Syst. Softw. Eng. 10 (4).
14. Kotonya, G., Sommerville, I., 1998. Requirements engineering with viewpoints. Software Engineering Journal, v.11.
15. Kurapati, N., Manyam, V.S.C., Petersen, K., 2012. Agile software development practice adoption survey. Agile Process. Softw. Eng. Entreme Programm.
16. Medeiros, J.R.V., 2017. Na approach to support the requirements specification in agile software development.
17. Nazir, N., Hasteer, N., Bansal, A., 2016. A survey on agile practices in the indian it industry. Proc. Of 6th INtl Conference on Coud System and Big Data Engineering (Confluence). IEEE.
18. Papatheocharous, E., Andreou, A.S., 2014. Empirical evidence and state of practice of software agile teams. J. Softw. 26 (9).
19. Rodríguez, P., Markkula, J., Oivo, M., Turula, K., 2012. Survey on agile and lean usage in finnish software industry. Proc. Of Intl Symposium on Empirical Software Engineering and Measurement (ESEM). ACM-IEEE.
20. Saito, S., Takeuchi, M., Hiraoka, M., Kitani, T., 2013. Requirements clinic: Third party inspection methodology and practice for improving the quality of software requirements specifications. Requirements Engineering Conference (RE), 21st IEEE International.
21. Salo, O., Abrahamsson, P., 2008. Agile methods in european embedded software development organisations: a survey on the actual use and usefulness of extreme programming and scrum. IET Softw. 2 (1).
22. Solinski, A., Petersen, K., 2016. Prioritizing agile benefits and limitations in relation to practice usage. Softw. Qual. J. 24 (2).
23. Sommerville, I., Software Engineering. Addison Wesley. 7th Ed.
24. Wang, X., Chao, L., Wang, Y., Sun, J., 2014. The role of requirements engineering practices in agile development: na empirical study. Proc. of 1st Asia Pacific Requirements Engineering Symposium, APRES Springer-Verlag Berlin Heidelberg.

25. Wohlin, C., Runeson, P., Höst, M., Ohlsson, M. C., Regnell, B., Wesslén, A. Experimentation in Software Engineering. Springer (2012).